\documentclass[12pt]{article}

\usepackage{graphicx}

\newcommand{\beq}{\begin{equation}}
\newcommand{\eeq}{\end{equation}}
\newcommand{\beqa}{\begin{eqnarray}}
\newcommand{\eeqa}{\end{eqnarray}}
\newcommand{\beqar}{\begin{eqnarray*}}
\newcommand{\eeqar}{\end{eqnarray*}}

\begin{document}
\thispagestyle{empty}

\vspace{32pt}
\begin{center}
\centerline{\textbf{\Large
Extra Higgs bosons in $t \bar t$ production at the LHC}}

\vspace{40pt}

Roberto Barcel\'o and Manuel Masip
\vspace{12pt}

\textit{
CAFPE and Departamento de F{\'\i}sica Te\'orica y del
Cosmos}\\ \textit{Universidad de Granada, E-18071, Granada, Spain}\\
\vspace{16pt}
\texttt{rbarcelo@ugr.es, masip@ugr.es}
\end{center}

\vspace{40pt}

\date{\today}

\begin{abstract}

The top quark has a large Yukawa coupling with the 
Higgs boson. In the usual extensions of the standard model the
Higgs sector includes extra scalars, which also tend to couple 
strongly with the top quark. Unlike the Higgs, these fields 
have a {\it natural} mass above $2m_t$, so they could introduce
anomalies in $t \bar t$ production at the LHC. We study their
effect on the $t \bar t$ invariant mass distribution 
at $\sqrt{s}=7$ TeV. We focus on the bosons ($H$,$A$) of
the minimal SUSY model and on the scalar field ($r$) associated to
the new scale $f$ in Little Higgs (LH) models. We show that in all 
cases the interference with the standard amplitude dominates over the 
narrow-width contribution. As a consequence, the
mass difference between $H$ and $A$ or the
contribution of an extra $T$-quark loop in LH models become
important effects in order to determine if 
these fields are observable there. We find that a 1 fb$^{-1}$
luminosity could probe the region $\tan\beta \le 3$ of SUSY
and $v/(\sqrt{2}f) \ge 0.3$ in LH models.

\end{abstract}

\newpage

\section{Introduction}

The main objective of the LHC is to reveal the nature of the
mechanism breaking the electroweak (EW) symmetry. This requires
not only a determination of the Higgs mass and couplings,
but also a search for additional particles that may be 
related to new dynamics or symmetries present
at the TeV scale. 
The top-quark sector appears then as a promising place 
to start the search, as it is there where the EW symmetry 
is broken the most (it contains the heaviest fermion).
Generically, the large top-quark 
Yukawa coupling with the Higgs boson ($h$) also 
implies large couplings with the extra physics. For example, in
SUSY extensions $h$ comes together with neutral scalar ($H$) 
and pseudoscalar ($A$) fields \cite{Djouadi:2005gj}. 
Or in Little Higgs (LH) models, a global symmetry in the Higgs 
and the top-quark sectors introduces
a scalar singlet and an extra $T$ quark 
\cite{Schmaltz:2005ky,Perelstein:2005ka}. In all cases these 
scalar fields have large Yukawa couplings that could
imply a sizeable production rate in hadron collisions and a
dominant decay channel into $t\bar t$. 

The energy and the luminosity to be achieved at the LHC make
this collider a top-quark factory, with around $1.5\times 10^5$ 
pairs at $\sqrt{s}=7$ TeV and 1 fb$^{-1}$. In this paper we 
study the possibility that the production and decay of 
extra Higgses distorts the $t\bar t$ invariant mass
distribution ($m_{t\bar t}$). The relevant amplitudes
are pictured in Fig.~1. We first review 
\cite{Gaemers:1984sj,Dicus:1994bm} the (analytical)
expressions for the cross section when the intermediate
field is a scalar or a pseudoscalar field and the loop
fermion is the top or a heavier $T$ quark. 
Then we define the models to be analyzed and study the
parton-level cross section in each case. Finally we discuss
the possible signal at the LHC.

\begin{figure}[b]
\begin{center}
\includegraphics[width=1\linewidth]{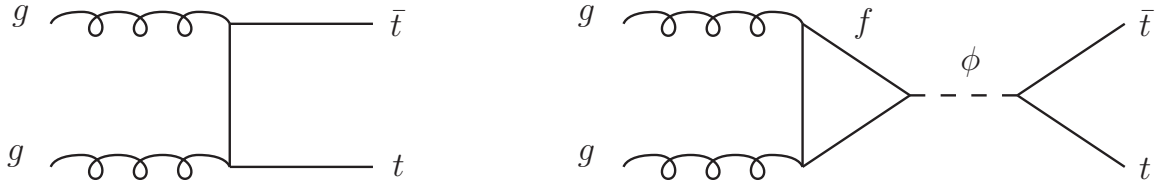} 
\end{center}
\caption{Diagrams that interfere in $t\bar t$ production.
\label{fig1}}
\end{figure}

\section{Top quarks from scalar Higgs bosons}

The potential to observe new physics in $m_{t\bar t}$
at hadron colliders has been discussed in 
previous literature 
\cite{Dicus:1994bm,Frederix:2007gi,Barger:2006hm,
:2007dia,Abazov:2008ny,Cabrera:2009zza,Baur:2007ck,
Hioki:2009hm,Kumar:2009vs}. 
In general, any heavy $s$--channel 
resonance with a significant
branching ratio to $t\bar t$ will introduce distortions:
a {\it bump} that can be evaluated in the narrow-width
approximation or more complex structures (a {\it peak} followed
by a {\it dip}) when interference effects are 
important \cite{Berdine:2007uv}. In the diagram depicted
in Fig.~1 the intermediate scalar is produced 
at one loop \cite{Georgi:1977gs}, but 
the gauge and Yukawa couplings are all strong.

Let us first consider a scalar $\phi$ 
coupled to the top quark and (possibly)
to a heavy fermion.
The leading-order (LO) 
differential cross section for $gg\rightarrow t\bar t$
is then 
\beqa
{{\rm d} \sigma \over {\rm d} z} &=& 
{{\rm d} \sigma_{QCD} \over {\rm d} z}+ {\alpha_s^2 \;
y_{\phi \bar t t}^2\; s^2\; \beta^3\over 1536\; \pi^3}\;
\left| {N(s)\over s-m_\phi^2+i\; m_\phi\Gamma_\phi(s)} \right|^2\\
&& 
-{\alpha_s^2\; y_{\phi \bar t t}\; m_t\; 
\beta^3\over 48\sqrt{2}\; \pi}\; {1\over 1-\beta^2 z^2}\;
{\rm Re} \left[ {N(s)\over s-m_\phi^2+i\; m_\phi\Gamma_\phi(s)}
\right] \;,
\label{csphi}
\eeqa
where $z=\cos\theta$ is the cosine of the angle 
between an incoming $g$ and
$t$, $m_t$ and $y_{\phi\bar t t}$ are the top-quark mass and 
Yukawa coupling, 
and $\beta=\sqrt{1-4 m_t^2/s}$
is the velocity of $t$ in the center of mass frame.
The function $N(s)$ associated to the fermion loop is
\beq
N(s)=\sum_f {3\; m_f\; y_{\phi \bar f f}\over \sqrt{2}\;s}
\left[ 1+\left( 1-{4m_f^2\over s} \right) I_f(s) \right]\;,
\eeq
where $f$ may be the top or another quark strongly coupled
to $\phi$, and 
$I_f(s)$ takes a different
form depending on the mass $m_f$:
\beq
I_f(s) = \left\{
\begin{array}{l l} 
\displaystyle
\left( {\rm Arcsin}\sqrt{s\over 4m_f^2} \right)^2 
& s < 4m_f^2\;; \\
\\
\displaystyle
-{1\over 4} \left( \ln {1+\sqrt{1-4 m_f^2/s}\over
1-\sqrt{1-4 m_f^2/s}} - i \;\pi \right)^2
& s > 4m_f^2\;.
\end{array} \right. 
\label{if}
\eeq
If $2m_f>\sqrt{s}$ then $I_f$ is real and the interference 
vanishes at $s=m_\phi^2$. If $f$ is the top or
any fermion with 
$2m_f<\sqrt{s}$, then this contribution can be seen as a 
final-state $f\bar f$ interaction \cite{Dicus:1994bm}. 
The diferential QCD contribution 
${\rm d} \sigma_{QCD}/ {\rm d} z$
can be found in \cite{Combridge:1978kx,Ellis:1986ef}.

For a pseudoscalar $A$ we have 
\beqa
{{\rm d} \sigma \over {\rm d} z} &=& 
{{\rm d} \sigma_{QCD} \over {\rm d} z}+ {3\;\alpha_s^2 \;
y_{A \bar t t}^2\; s^2\; \beta \over 512\; \pi^3}\;
\left| {P(s)\over s-m_A^2+i\; m_A\Gamma_A(s)} \right|^2\\
&& 
-{\alpha_s^2\; y_{A \bar t t}\; m_t\; 
\beta \over 16\sqrt{2}\; \pi}\; {1\over 1-\beta^2 z^2}\;
{\rm Re} \left[ {P(s)\over s-m_A^2+i\; m_A\Gamma_A(s)}
\right] \;,
\label{csA}
\eeqa
with
\beq
P(s)=\sum_f {m_f\; y_{A \bar f f}\over \sqrt{2}\; s}
\;I_f(s)\;.
\eeq

As we will see in the next section, to have 
an observable effect it is essential that the width
$\Gamma_\phi$ is small. This is precisely the
reason why the effect on $m_{t\bar t}$
of a very heavy standard Higgs $h$ would
be irrelevant. A 500 GeV Higgs boson would couple strongly
to the top quark, but even stronger to itself: 
$\lambda=m_h^2/(2v^2)\approx 2$.
Its decay into would-be Goldstone bosons (eaten by the 
massive $W$ and $Z$) would then dominate, implying a 
total decay width 
\beq
\Gamma_h \approx 
{3\over 8\pi\; v^2} \left[
{m_t^2 m_h\;\beta_t^3} 
+{m_h^3 \over 4} \left(
\beta_V^3 +{3\over 4} \beta_V(1-\beta_V^2)^2\right)
\right]\;\approx 60\; {\rm GeV}\;,
\eeq
where 
\beq
\beta_{t(V)}=\sqrt{1-{4m_{t(V)}^2\over m_h^2}}\;
\eeq
and we have taken a common $W,Z$ mass $m_V\approx 90$ GeV.

\begin{figure}
\begin{center}
\includegraphics[width=0.45\linewidth]{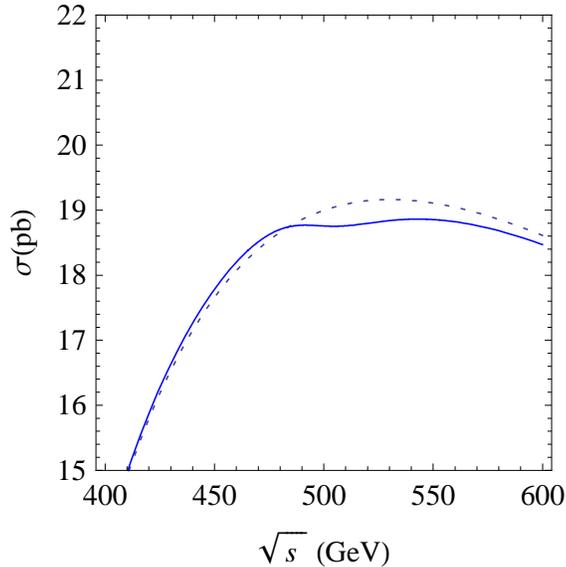} 
\end{center}
\caption{$\sigma(gg\rightarrow t\bar t)$
with a standard Higgs of mass $m_h=500$ GeV.
\label{fig2}}
\end{figure}
The plot 
in Fig.~2 shows a too small deviation due to the standard Higgs
in $\sigma(gg\rightarrow t\bar t)$. 
To have a smaller width and a larger effect
the mass of the resonance must {\it not} be EW.
In particular, SUSY or LH models
provide a new scale and massive Higgses with no need for large
scalar self-couplings.

\section{SUSY neutral bosons}
SUSY incorporates two Higgs doublets, and after EW
symmetry breaking there are two neutral bosons ($H$ and
$A$) in addition to the light Higgs. The mass of these two fields 
is not EW (it comes from the SUSY breaking
sector), so they are {\it naturally} heavy enough to decay in
$t \bar t$. Their mass difference is EW,
of order $M_Z^2/m_A$, with important top-quark corrections
at the loop level. More precisely, the relation between $m_A$ 
and $m_{H,h}$ is \cite{deBoer:1994he}
\beq
m_{H,h}^2={1\over 2} \left( m_A^2+M_Z^2+\Delta_{11}+
\Delta_{22}\pm \sqrt{\Delta_0^2} \right)
\eeq
where
\beqa
\Delta_0^2 &=&\left( m_A^2+M_Z^2+\Delta_{11}+
\Delta_{22} \right)^2-4\;m_A^2 M_Z^2 \cos^22\beta 
\nonumber \\
&&-4 \left( \Delta_{11} \Delta_{22}- \Delta_{12}^2\right)
-4 \left( M_Z^2\cos^2\beta +m_A^2\sin^2\beta \right)\Delta_{22}
\nonumber  \\
&&-4 \left( M_Z^2\sin^2\beta +m_A^2\cos^2\beta \right) \Delta_{11} 
-4 \;\sin 2\beta \left( M_Z^2+m_A^2 \right) \Delta_{12} \;,\\
%
\Delta_{11}&=&{3g^2\over 16\pi^2}\; {m_t^4\over M_W^2 \sin^2\beta}
\left[{\mu\left( A_t m_0-\mu \cot\beta\right)\over 
\tilde m_{t1}^2-\tilde m_{t2}^2}\right]^2
d(\tilde m_{t1}^2,\tilde m_{t2}^2)\;,\\
%
\Delta_{22}&=&{3g^2\over 16\pi^2}\; {m_t^4\over M_W^2 \sin^2\beta}
\;\Bigg[\; {2 A_t m_0 \left( A_t m_0-\mu \cot\beta\right)\over 
\tilde m_{t1}^2-\tilde m_{t2}^2}\;
\ln {\tilde m_{t1}^2\over \tilde m_{t2}^2}
\nonumber \\
&&+\ln {\tilde m_{t1}^2 \tilde m_{t2}^2\over m_t^4}
+\left({ A_t m_0 \left( A_t m_0-\mu \cot\beta\right)\over 
\tilde m_{t1}^2-\tilde m_{t2}^2}\right)^2
d(\tilde m_{t1}^2,\tilde m_{t2}^2)\;\Bigg]\;,\\
%
\Delta_{12}&=&-{3g^2\over 16\pi^2}\; {m_t^4\over M_W^2 \sin^2\beta}
\;{\mu\left( A_t m_0-\mu \cot\beta\right)\over 
\tilde m_{t1}^2-\tilde m_{t2}^2}\;
\Bigg[\; \ln {\tilde m_{t1}^2\over \tilde m_{t2}^2}
\nonumber \\
&&+{ A_t m_0 \left( A_t m_0-\mu \cot\beta\right)\over 
\tilde m_{t1}^2-\tilde m_{t2}^2} \;
d(\tilde m_{t1}^2,\tilde m_{t2}^2)\;\Bigg]\;, 
\eeqa
and 
\beq
d(m_{1}^2, m_{2}^2)=2-{m_1^2+m_2^2 \over
m_1^2-m_2^2}\;\ln {m_1^2\over m_2^2}\;.
\eeq
Varying the $\mu$ parameter and the stop masses and trilinears, 
for $m_A= 500$ GeV
we obtain typical values of $m_H-m_A$ between $-2$ and $+10$ GeV.

The scalar masses of interest correspond to the {\it decoupling}
regime, where $h$ is basically the SM Higgs and 
\beq
y_{H\bar t t}\approx -{m_t\sqrt{2}\over v}\;{1\over \tan\beta}\approx
-y_{A\bar t t}\;.
\eeq
In addition, we will consider low values of $\tan\beta$, where
the decay into bottom quarks is not important and the 
(energy-dependent) widths can be approximated to
\beq
\Gamma_H(s)\approx {3\;y_{H\bar t t}^2\; s\;\beta^3\over 16\pi\; m_H}\;,
\;\;\;
\Gamma_A(s)\approx {3\;y_{A\bar t t}^2\; s\;\beta\over 16\pi\; m_A}\;.
\eeq

\begin{figure}
\begin{center}
\begin{tabular}{ccc}
\includegraphics[width=0.45\linewidth]{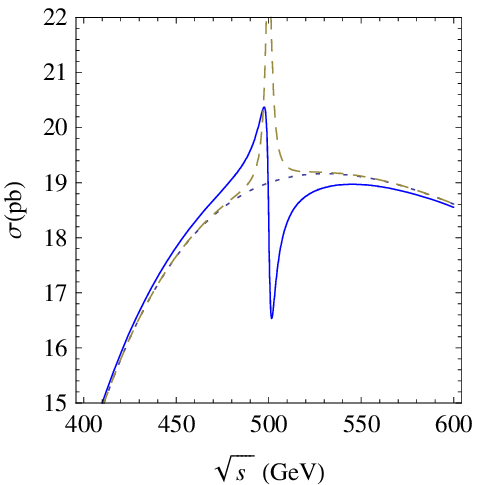} & $\;\;\;$ &
\includegraphics[width=0.45\linewidth]{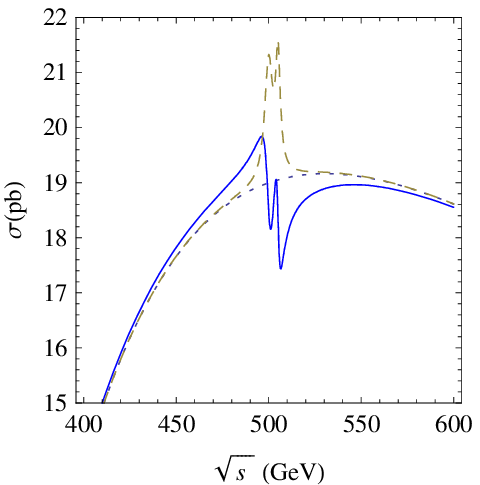} 
\end{tabular}
\end{center}
\caption{$\sigma(gg\rightarrow t\bar t)$ for $\tan\beta=2$
and SUSY bosons of mass $m_A=m_H=500$ GeV (left)
or $m_A=500$, $m_H=505$ GeV (right). Dashes provide 
the narrow-width approximation and dots the standard model
cross section.
\label{fig3}}
\end{figure}

\begin{figure}
\begin{center}
\includegraphics[width=0.43\linewidth]{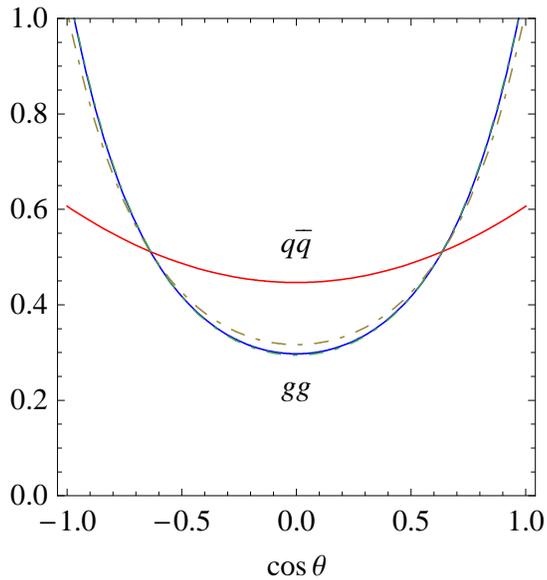} 
\end{center}
\caption{Standard angular distribution for the $t$ quarks
from $q\bar q$ and $g g$ collisions 
at $\sqrt{s}=500$ GeV. We include (dashes) the distribution
from $g g$ at the peak and the dip of Fig.~3--left.
\label{fig4}}
\end{figure}

In Fig.~3 we plot $\sigma(gg\rightarrow t\bar t)$ at center of
mass energies around $m_A=500$ GeV for  
$m_H=500$ GeV (left) and $m_H=505$ GeV (right). We have
taken $\tan\beta=2$, which implies $\Gamma_H\approx 3.0$ GeV 
and $\Gamma_A\approx 5.3$ GeV.
We observe an average 5.5\% excess and  
8.1\% deficit in the 5 GeV intervals before and after 
$\sqrt{s}=500$ GeV, respectively. We include in dashes
the result ignoring the interference (last term in 
Eqs.~(\ref{csphi}) and (\ref{csA})), which would not be captured if one
uses the narrow-width approximation. It is apparent that the 
interference with the standard amplitude gives the dominant effect. 
In Fig.~3--left the position of
the peaks and dips caused by
$H$ and $A$ overlap {\it constructively}
(notice, however, that in this $CP$ conserving Higgs sector 
their amplitudes do not interfere). In contrast, in  Fig.~3--right
their mass difference implies a partial cancellation between 
the dip caused by $A$ and
the peak of $H$. 

The scalar and pseudoscalar couplings with the top quark grow
at smaller values of $\tan\beta$, increasing 
the cross section and the scalar width.
For example, for $\tan\beta=1$ the excess at $\sqrt{s}<500$ GeV
grows to the 6.2\% and 
the deficit to the 9.7\%, whereas for $\tan\beta=5$ the excess 
and deficit
are just a 2.1\% and a 2.6\%, respectively.

The normalized 
angular distribution of the $t$ quark in the center of mass
frame is given in Fig.~4. We plot the standard distributions for
top-quark production in $g g$ and $q\bar q$ collisions
together with the distribution from $gg$ at the peak and the 
dip obtained in Fig.~3--left. In the narrow-width 
approximation a scalar resonance gives a flat contribution. However, we
find that the excess or deficit from the scalar interference 
\begin{figure}[h]
\begin{center}
\begin{tabular}{ccc}
\includegraphics[width=0.45\linewidth]{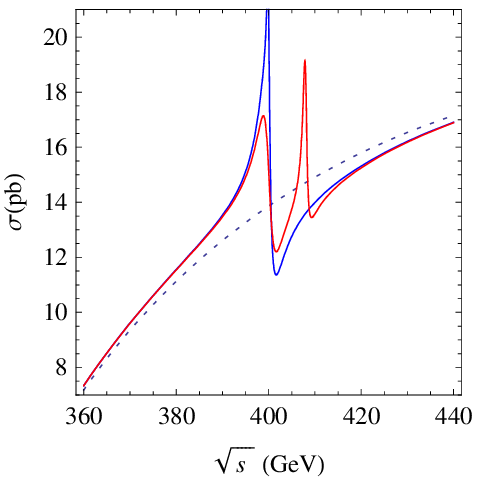}  & $\;\;\;$ &
\includegraphics[width=0.46\linewidth]{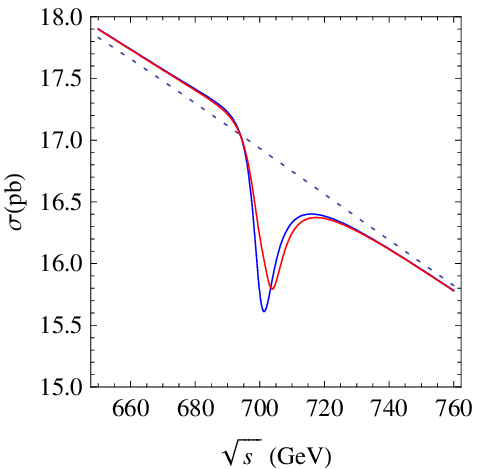} 
\end{tabular}
\end{center}
\caption{$\sigma(gg\rightarrow t\bar t)$ for $m_A=400$ GeV
and $m_A=700$ GeV. We include the cases 
$m_H=400,408$ GeV (left) and $m_H=700,703$ GeV (right) 
\label{fig5}}
\end{figure}
is {\it not} flat and 
does not change significantly the angular distribution. Different
cuts could be applied to reduce the background for $t\bar t$ 
production at the LHC \cite{Barger:2006hm} or even to optimize 
the contribution from $gg$ versus $q\bar q$, but not to enhance
the relative effect of the scalars on $\sigma(gg\rightarrow t\bar t)$.

In Fig.~5 we plot the parton cross section for lower and higher
values of the pseudoscalar mass ($m_A=400, 700$ GeV).
We include the cases where the boson $H$ is degenerate with
$A$ or slightly heavier ($m_H=408$ GeV and $m_H=703$ GeV). We see that 
at lower scalar masses the peak dominates, whereas for
large values of $m_A$ the dip is the dominant effect. This
behaviour, related to the slope of the standard cross section,
reduces in both cases the relevance of the mass difference between the
scalar and the pseudoscalar Higgses.

\section{Little Higgs boson}
In LH models the Higgs appears as a pseudo-Goldstone
boson of a global symmetry broken spontaneously 
at the scale $f>v/\sqrt{2}=174$ GeV. The global 
symmetry introduces an extra $T$ quark
that cancels top-quark quadratic corrections 
to the Higgs mass parameter. 
The presence of this vectorlike $T$ quark and of a massive 
scalar singlet (the {\it Higgs} 
of the symmetry broken at $f$) 
are then generic features in all these models.

Once the electroweak VEV is included the 
doublet and singlet Higgses (and also the $t$ and $T$
quarks) mix \cite{Barcelo:2008je,Barcelo:2007if}. 
The singlet component $\approx v/(\sqrt{2} f)$
in $h$ will reduce its coupling both to the
top quark and to the gauge bosons and, in turn, 
$r$ will get a doublet
component that couples to these fields.

\begin{figure}
\begin{center}
\begin{tabular}{ccc}
\includegraphics[width=0.33\linewidth]{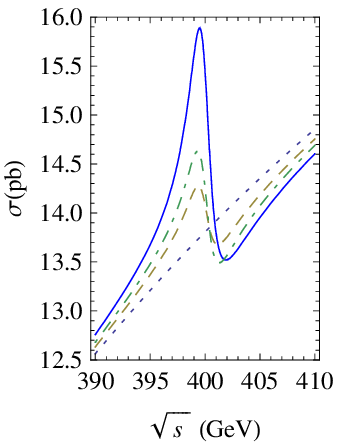} & 
\includegraphics[width=0.32\linewidth]{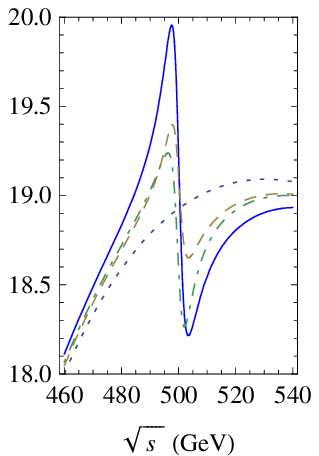} & 
\includegraphics[width=0.32\linewidth]{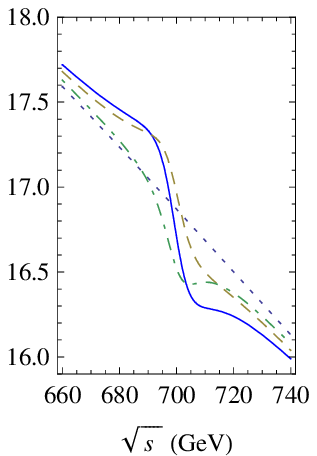} 
\end{tabular}
\end{center}
\caption{$\sigma(gg\rightarrow t\bar t)$ for a LH model with 
$s_\theta=0.5$, $m_r=500$ GeV and $m_T=400,500,700$ GeV. 
Dashes (dot-dashes) correspond to an amplitude with only the 
$T$ ($t$) quark loop.
\label{fig6}}
\end{figure}

It is easy to see that the most general\footnote{There could
be an additional mixing, $T\rightarrow c_\beta T + s_\beta t$ 
in the second line of Eq.~(\ref{yt}), but it must be small 
\cite{Aguilar-Saavedra:2002kr} to avoid a too large value of $V_{Tb}$.}
top-quark Yukawa sector with no quadratic corrections at
one loop is 
\beqa
-{\cal L}_t &=& \lambda \; \left( f+{r\over \sqrt{2}}\right)
\sin{u+h\over\sqrt{2}f} \left(c_\alpha t + s_\alpha T \right) t^c
\nonumber \\
&+& \lambda \; \left( f+{r\over \sqrt{2}}\right)
\cos{u+h\over \sqrt{2}f} \; T T^c
+ {\rm h.c.}\;,\label{yt}
\eeqa
where $u$ and $f$ are VEVs satisfying
\beq
f\; \sin {u\over\sqrt{2}f} \equiv f\; s_\theta = {v\over \sqrt{2}}\;.
\eeq
Eq.~(\ref{yt}) becomes 
\beqa
-{\cal L}_t &=& \lambda \; \left( f+{r\over \sqrt{2}}\right)
\left( s_\theta \cos {h\over\sqrt{2}f} +  
c_\theta \sin{h\over\sqrt{2}f} \right)
 \left(c_\alpha t + s_\alpha T \right) t^c +
\nonumber \\
&& \lambda \; \left( f+{r\over \sqrt{2}}\right)
\left( c_\theta \cos {h\over\sqrt{2}f} - 
s_\theta \sin{h\over\sqrt{2}f} \right) T T^c
+ {\rm h.c.}\;.\label{yt2}
\eeqa
Fermion masses, Yukawa couplings and dimension-5 operators
(necessary to check the cancellation of all one-loop 
quadratic corrections) are then obtained by expanding 
\beq
\cos {h\over\sqrt{2}f}\approx 1-{h^2\over 4 f^2}\;,\;\;\;
\sin {h\over\sqrt{2}f}\approx {h\over \sqrt{2} f}
\eeq
The fermion masses and the Yukawas to the heavier scalar 
$r$ have the same structure,
\beq
-{\cal L}_t\supset \lambda \; \left( f+{r\over \sqrt{2}}\right)
\left(\begin{array}{cc} t & T \end{array}\right)
\left(\begin{array}{cc} s_\theta c_\alpha & 0 \\
s_\theta s_\alpha &  c_\theta^2 \end{array}\right)
\left(\begin{array}{c} t^c \\ T^c \end{array}\right)\;.
\eeq
This implies 
\beq 
y_{r\bar t t}={m_t\over f}={\sqrt{2}\; s_\theta\; m_t\over v}\;\;\;
{\rm and}\;\;\;
y_{r\bar T T}={m_T\over f}\;,
\eeq
where the quarks are mass eigenstates. 
The mass of the heavier $T$ quark is 
$m_T\approx m_t c_\theta/(s_\theta c_\alpha)$, and its mixing
with the doublet 
$V_{Tb}\approx s_\theta^2 s_\alpha c_\alpha/c_\theta^2$.

The extra Higgs $r$ is somehow similar to the heavier scalar 
in a doublet plus singlet model, with the 
doublet component growing with $s_\theta=v/(\sqrt{2} f)$.
If $s_\theta$ is sizeable so is its coupling to the top
quark. The coupling to the extra $T$ quark is stronger, 
but if $r$ is lighter
than $2m_T$ then its main decay mode will be into $t\bar t$.
Actually, the doublet component in $r$ may also imply large
couplings to the would-be Goldstone bosons for large values
of $m_r$. More precisely, its decay width $\Gamma_r(s)$ 
at $4m_t^2<s<4m_T^2$ is
\beq
\Gamma_r(s)\approx  
{3\;s_\theta^2\; s\over 8\pi\; v^2} \left[
{m_t^2\;\beta_t^3 \over m_r} 
+{s_\theta^2\; m_r \over 4} \left(
\beta_V^3 +{3\over 4} \beta_V(1-\beta_V^2)^2\right)\;.
\right]\;
\eeq
Therefore, $r$ is a naturally heavy ($m_r\approx f$) 
but narrow scalar
resonance with large couplings to quarks and an order one
braching ratio to $t \bar t$.

In Fig.~6 we plot the parton cross section 
$\sigma(gg\rightarrow t\bar t)$ for $s_\theta=0.5$, $m_T=500$ GeV
and several values of $m_r$. 
We separate the contributions from the top
and the $T$ quark loops (the second one vanishes at $s=m_r^2$).
The plot is similar to the one obtained for SUSY bosons of
the same mass.
At higher values of $m_r$ the decay width 
$\Gamma_{r}$ grows, diluting the 
effect (see Fig.6, right). In
contrast, for lower masses the scalar $r$ has a 
narrow width and is strongly coupled 
to quarks, which produces a larger effect (in Fig.6, left).
The contribution from the standard $t$-quark loop grows with
$s_\theta$, whereas the contribution from the
extra $T$-quark is 
basically independent of $m_T$.

\section{Signal at the LHC}

Let us now estimate the invariant mass distribution 
of $t\bar t$ events ($m_{t\bar t}$) in $pp$ collisions at the LHC. 
To evaluate the hadronic cross sections we 
will use the MSTW2008 PDFs \cite{Martin:2009iq}. 
The effect of next-to-leading order (NLO) corrections on the
expressions given in previous sections has been 
studied by several groups (see for example 
\cite{Frederix:2007gi,Frixione:2003ei}).
In particular, the authors in \cite{Frederix:2007gi} analyze 
the dependence of ${\rm d}\sigma/{\rm d} m_{t\bar t}$ on the choice of 
renormalization and factorization scales and of PDFs. They 
show that if the LO cross section 
is normalized to the NLO one at low values of 
$m_{t\bar t}$, 
then the deviations introduced by these scales and by
the uncertainty in the PDFs at $m_{t\bar t}<1$ TeV 
are small (order 10\%). For (scalar and
pseudoscalar) Higgs production in $pp$
collisions and Higgs decay, a complete 
review of NLO results can be found in \cite{Djouadi:2005gj}. 
From the expressions there we obtain that QCD corrections 
enhance the production cross section  
in approximately 
a 20\%, and that the Higgs decay width into $t\bar t$ (for 
$m_{\phi}\gg 2m_t$) is also increased in around a 10\%.
Given these estimates, we have evaluated $pp\rightarrow t\bar t$ 
taking fixed renormalization and factorization 
scales ($\mu_{R,F}=m_t$) and normalizing the LO result 
to the NLO cross section in \cite{Frederix:2007gi} with a 
global factor of 1.3. Our differential cross section 
coincides then with that NLO result at $m_{t\bar t}=500$ GeV.

\begin{figure}
\begin{center}
\begin{tabular}{ccc}
\includegraphics[width=0.33\linewidth]{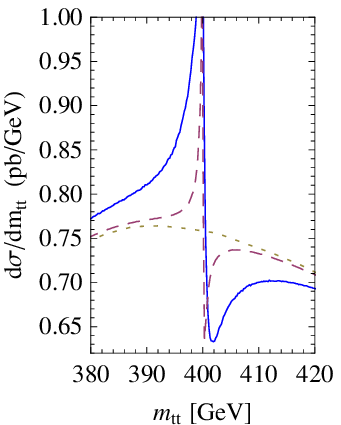} & 
\includegraphics[width=0.318\linewidth]{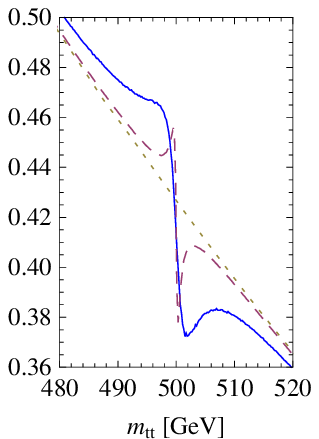} & 
\includegraphics[width=0.328\linewidth]{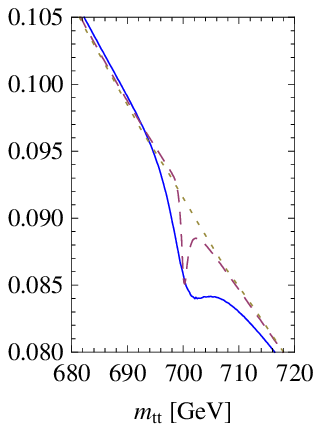} 
\end{tabular}
\end{center}
\caption{
${\rm d}\sigma/{\rm d} m_{t\bar t}$ in SUSY models with
$m_H=m_A=400,500,700$ GeV and $\tan\beta=2$ (solid),
$5$ (dashes). 
\label{fig7}}
\end{figure}

\begin{figure}
\begin{center}
\begin{tabular}{ccc}
\includegraphics[width=0.33\linewidth]{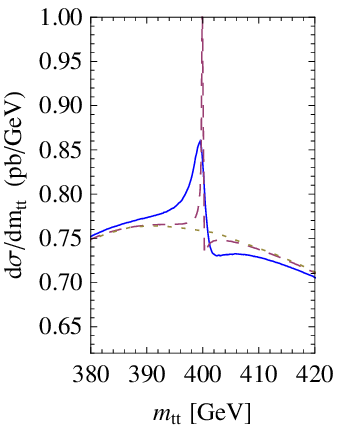} & 
\includegraphics[width=0.318\linewidth]{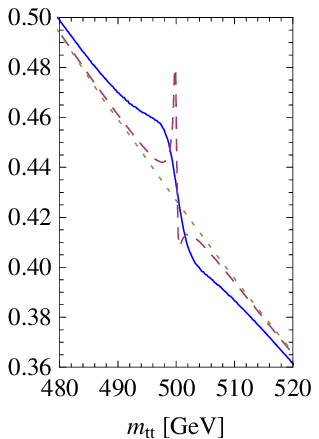}  & 
\includegraphics[width=0.328\linewidth]{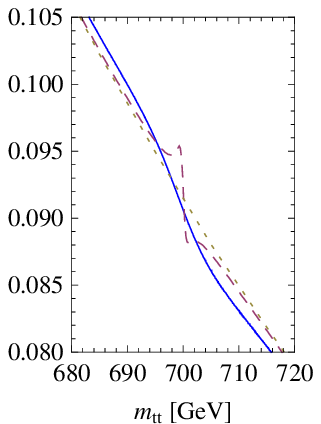} 
\end{tabular}
\end{center}
\caption{
${\rm d}\sigma/{\rm d} m_{t\bar t}$ in LH models with 
$m_r=400,500,700$ GeV and $s_\theta=0.5$ (solid),
$0.2$ (dashes).
\label{fig8}}
\end{figure}

\begin{figure}
\begin{center}
\includegraphics[width=0.47\linewidth]{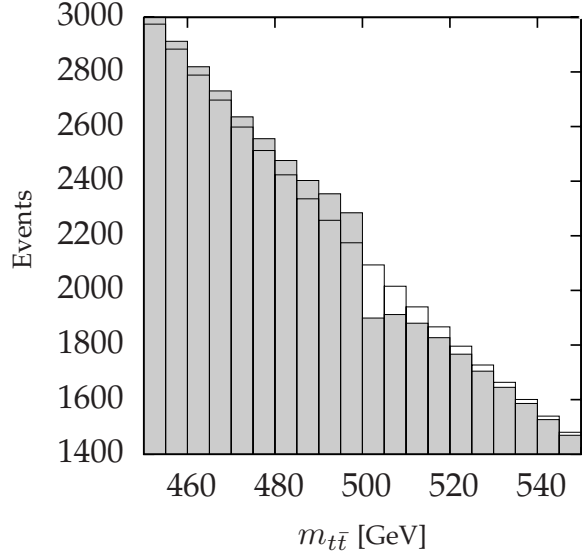} 
\end{center}
\caption{Number of $t\bar t$ events in $pp$ collisions
at 7 TeV and 1 fb$^{-1}$ for
$m_A=m_H=500$ GeV and $\tan\beta=2$ distributed in 5 GeV bins.
\label{fig9}}
\end{figure}

\begin{figure}
\begin{center}
\begin{tabular}{ccc}
\includegraphics[width=0.47\linewidth]{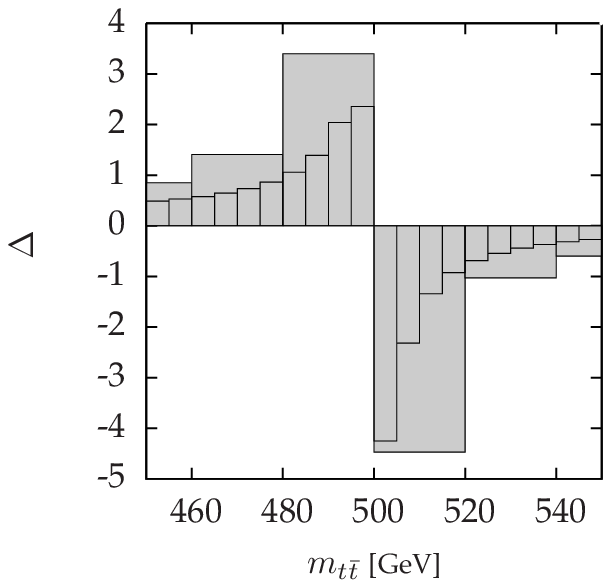} & $\;\;\;$ &
\includegraphics[width=0.47\linewidth]{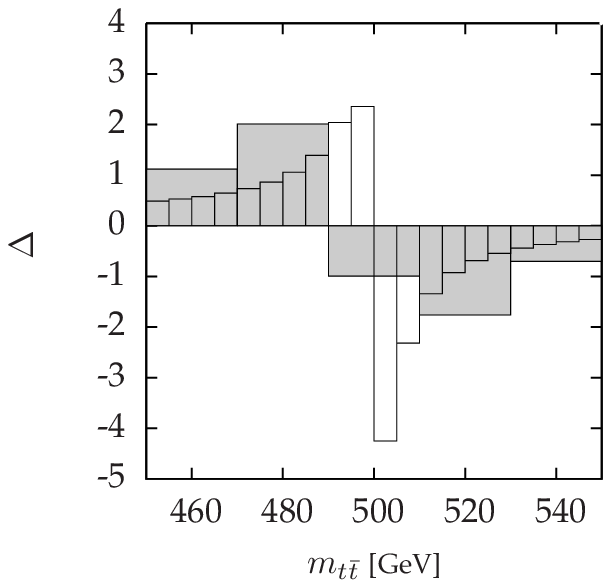} 
\end{tabular}
\end{center}
\caption{Deviation $\Delta=(N-N_{SM})/\sqrt{N_{SM}}$ in the 
number of events 
respect to the standard prediction for two
different binning ($m_A=m_H=500$ GeV and $\tan\beta=2$). 
\label{fig10}}
\end{figure}

We will take a center of mass energy of 7 TeV. We obtain
that at these energies the cross section $pp\rightarrow t\bar t$
is dominated by $gg$ fusion, with 
$q\bar q\rightarrow t\bar t$ accounting for just 12\% of the
top-quark pairs. In Figs.~7,8 we plot
${\rm d}\sigma/{\rm d} m_{t\bar t}$ 
for some of the SUSY and LH models described in Sections
3-4.  These figures {\it translate} 
the parton-level cross sections in Figs.~3--6 into anomalies 
in the invariant mass distribution in $pp$ collisions.

To estimate the possible relevance at the LHC of these 
cross sections, 
we will calculate the number of $t\bar t$ events
assuming an integrated 
luminosity of 1 fb$^{-1}$ (expected at the 7 TeV phase
of the collider), and we will not apply any cuts. 
In Fig.~9 we plot the number of events per 5 GeV bin of
$m_{t\bar t}$ in
the SUSY model with $m_A=m_H=500$ GeV and $\tan\beta=2$.
We observe a 5\% excess followed by a 9\% deficit, with
smaller deviations as $m_{t\bar t}$ separates from the mass
of the extra Higgs bosons. In Fig.~10 we distribute the 
events in 20 GeV bins and plot the statistical significance 
\beq
\Delta \equiv {N-N_{SM}\over \sqrt{N_{SM}}}
\eeq
of the deviations, where $N$ is the total number
of events in the bin.
The typical signal is an increasing excess in 
a couple of 20 GeV bins that may reach a $+3.4\sigma$ deviation
followed by a deficit of $-4.5\sigma$.
We find that changing the binning is important in order to
optimize the effect. If the same 20 GeV bin includes the peak and
the dip (Fig.~10, right) then the maximum deviation is just a
$\pm 2\sigma$ effect.

The result is very similar for a LH scalar of
$m_r=500$ GeV with $s_\theta=0.5$. In this LH model
we obtain deviations in consecutive 20 GeV bins 
reaching $+2.5\sigma$ and $-2\sigma$. However, the effect is a bit
more localized, and the cancellation if peak and dip coincide
in a bin is stronger: it may result in three bins 
with just $+1.3\sigma$, $+0.6\sigma$ and $-1.2\sigma$ deviations.

The binning is less important for larger Higgs
masses. For example, in the SUSY case with $m_A=m_H=700$ GeV 
the typical sequence is a couple of 20 GeV bins with a slight
$+0.2\sigma$ excess followed by $-1.2\sigma$, $-0.4\sigma$ and 
$-0.2\sigma$ deficits. In the LH model with $m_r=700$ GeV
the initial excess (caused by the $T$-quark loop)
is a bit more significant, a typical sequence
would consist of two bins with $+0.4\sigma$ excess followed by 
$-0.8\sigma$ and $-0.4\sigma$ deficits.

Let us finally focus on lighter Higgses, as they provide the
most promising signal. In Fig.~11 we plot the event distribution
(left) and the statistical significance (right) 
for $\tan\beta=2$ and $m_A=m_H=400$ GeV,
whereas Fig.~12 corresponds to a mass difference of 8 GeV
($m_A=400$ GeV and $m_H=408$ GeV). The sequence of deviations in
both cases would be seen as a clear anomaly, reaching an excess
of up to $13\sigma$ (for $m_H-m_A=-2$ GeV) in a 20 GeV bin. 
The LH case
is analogous but, again, more localized. We obtain an excess 
of $+3.4\sigma$ in a 20 GeV bin followed by a $-1.7\sigma$ deficit.

\begin{figure}
\begin{center}
\begin{tabular}{ccc}
\includegraphics[width=0.47\linewidth]{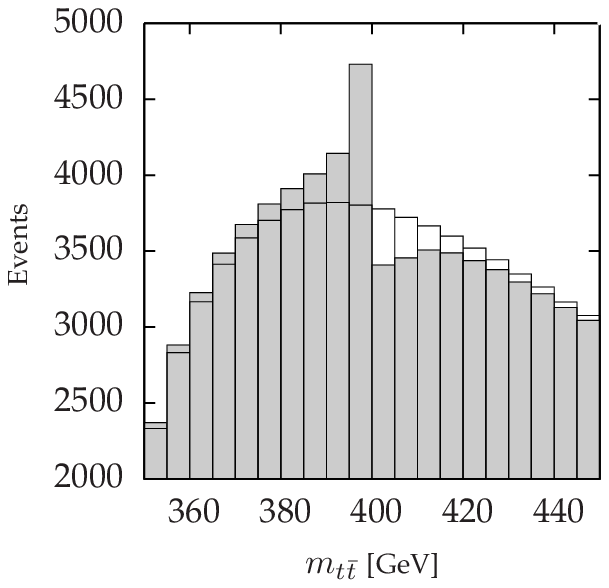} & $\;\;\;$ &
\includegraphics[width=0.47\linewidth]{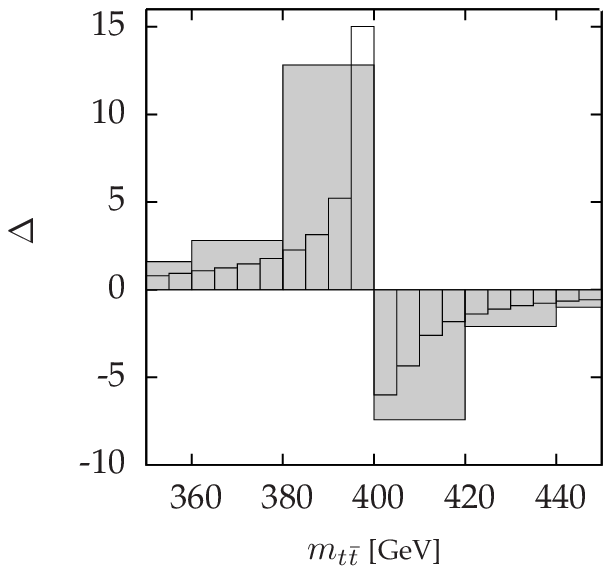} 
\end{tabular}
\end{center}
\caption{Number of $t\bar t$ events in $pp$ collisions (left) 
and deviation $\Delta$ (right)
for $m_A=m_H=400$ GeV and $\tan\beta=2$.
\label{fig11}}
\end{figure}

\begin{figure}
\begin{center}
\begin{tabular}{ccc}
\includegraphics[width=0.47\linewidth]{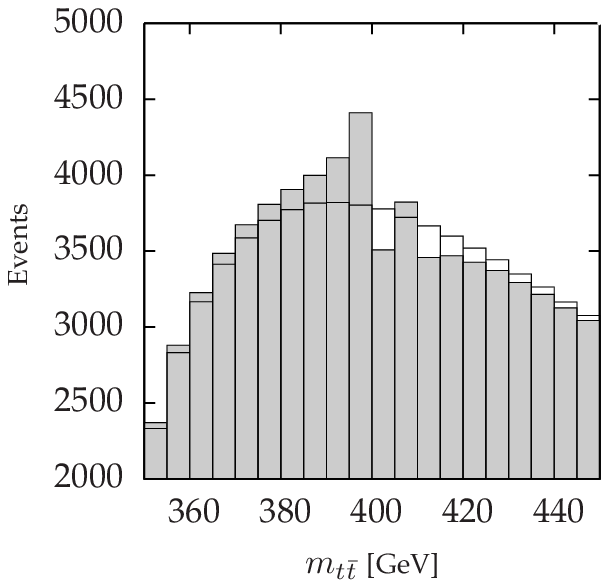} & $\;\;\;$ &
\includegraphics[width=0.47\linewidth]{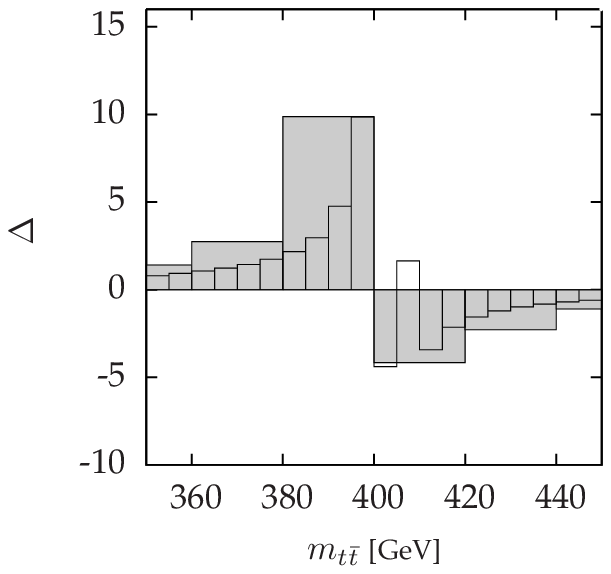} 
\end{tabular}
\end{center}
\caption{Number of $t\bar t$ events in $pp$ collisions (left) 
and deviation $\Delta$ (right)
for $m_A=400$ GeV, $m_H=408$ GeV and $\tan\beta=2$.
\label{fig12}}
\end{figure}

\section{Summary and discussion}

In models with an extended Higgs sector the extra bosons
tend to have large couplings with the top quark that imply
a sizeable one-loop production rate at hadron colliders.
If the mass of these bosons is not EW but 
comes from a new scale 
({\it e.g.}, the SUSY or the global symmetry-breaking 
scales), then they may decay predominantly  
into $t\bar t$. We have studied their effect on the $t\bar t$
invariant mass distribution at 7 TeV and  
1 fb$^{-1}$. We have considered
the deviations due to the neutral bosons $A$ and $H$ of the MSSM,
and to the scalar $r$ associated to the scale $f$ in LH
models. In all cases the interference dominates, 
invalidating the narrow-width approximation. 

The effect for
masses around 500 GeV is a peak followed by a dip of similar
size. In the SUSY case, values of $m_H-m_A$ smaller 
than 3 GeV enhance the 
deviation, whereas for larger values the effects 
at $m_A<m_{t\bar t}<m_H$ tend to cancell each other.
For $\tan\beta=2$ the significance of the signal, that can be optimized by 
changing the binning, results in sequences of 20 GeV bins with 
$+1.4\sigma$, $+3.4\sigma$, $-4.5\sigma$, $-1.0\sigma$ deviations.
In LH models the field $r$ couples both to the top
and to an extra $T$ quark. The main difference with the SUSY case 
is that the $T$ quark is heavy and the one-loop form factor 
to produce the scalar does not get an imaginary part.
To have a significant effect the doublet component 
$s_\theta=v/(\sqrt{2}f)$ in $r$ must be large and the 
extra $T$ quark heavier than $m_r/2$ (to close the $T\bar T$
decay channel). As $s_\theta$ grows $r$ resembles the
standard model Higgs, but with a singlet component that reduces
its width. The signal at the LHC for $s_\theta=0.5$ and 
$m_r=500$ GeV is similar to the SUSY case just described.

At larger scalar masses the peak decreases and the effect is
basically a dip in the invariant mass distribution. For 
$m_A=700$ GeV and 
$\tan\beta=2$ we get a couple of 20 GeV bins with a deficit of 
$-1.2\sigma$, and $-0.4\sigma$. The effect that
one may expect in LH models is alike, although 
(due to the contribution of the $T$-quark loop) the difference
between peak and dip is smaller.

Lower scalar masses provide the most promising scenario. Here
the peak dominates both in SUSY and LH models. In the SUSY case
with $m_A=400$ GeV and $m_H=408$ GeV the sequence of 20 GeV bins
at 7 TeV and 1 pb$^{-1}$ consists of 
$+2.7\sigma$, $+9.9\sigma$, $-4.2\sigma$ and $-2.3\sigma$
deviations. The signal increases in up to a 30\% if the mass 
difference is smaller (the optimal case is $m_H-m_A=-2$ GeV).
In the LH case with $s_\theta=0.5$ and $m_r=400$ GeV the 
sequence reaches a $+3.4\sigma$ 
deviation. All the effects grow for lower values of $\tan\beta$
and of the LH scale $f$. Obviously, their observability will
be better if the LHC reaches 14 TeV and higher luminosities.

An important observation is that the angular distribution of
the $t$ quark is unaffected by the intermediate scalar or
pseudoscalar bosons. The excess or the deficit caused by its
interference with the standard amplitude does 
{\it not} have a flat distribution in the center of mass frame, 
as one obtains in the narrow-width approximation.

Finally, we would like to comment on the possibility to
observe this type of signal at the Tevatron, which may achieve
10 fb$^{-1}$ at $\sqrt{s}=1.96$ TeV. The main difference with
the LHC is that at the Tevatron 90\% of the top-quark pairs
are produced through $q\bar q$ interactions. Since the signal 
that we have explored is caused by interference in the 
$gg\rightarrow t\bar t$ channel, for the same integrated
luminosity the deviations there would be 
9 times weaker than at the LHC (where gluon fusion provides
90\% of the top pairs). We find, however, that 1$\sigma$ 
deviations could be obtained at the Tevatron 
for low masses of the heavy Higgs bosons.
This signal could be enhanced by {\it separating} the $t\bar t$
events in two or three sets according to the $\cos \theta$
of the final $t$ quark. As we see in Fig.~4, the $gg$ and
$q \bar q$ contributions at 
$m_{t\bar t}\approx m_{\phi} \approx 500$ GeV 
have different angular distributions
(this difference, however, vanishes at lower invariant 
masses). One could separate, for example, the events with 
$\left| \cos\theta \right|$
larger or smaller than 0.6. Then the
anomalies in $d\sigma/ d m_{t\bar t}$ that we have discussed
should increase in the $\left| \cos\theta \right| > 0.6$
interval.

Although the generic effect on the $t\bar t$ invariant mass
distribution caused by a scalar field
with strong couplings to the top quark is known, we think that 
it is also important to study particular models. We find that 
the peculiar anomalies that appear in these two extensions
of the standard model may
be optimized by changing the binning and by applying the same cuts
that select $t \bar t$ production from gluon fusion.

\section*{Acknowledgments}
We would like to thank Nuno Castro, Mark Jenkins 
and Olaf Kittel for valuable discussions.
This work has been partially supported by
MICINN of Spain (FPA2006-05294) and by Junta de Andaluc\'{\i}a
(FQM 101 and FQM 437).

\end{document}